\documentclass[conference, letter]{IEEEtran}
\IEEEoverridecommandlockouts
\usepackage{verse}

\usepackage{amsfonts}
\usepackage{caption}
\usepackage{subcaption}
\usepackage[bottom]{footmisc}
\usepackage{cite}
\usepackage{amsmath,amssymb,amsfonts}
\usepackage{lipsum}
\usepackage{graphicx}
\usepackage{textcomp}
\usepackage{xcolor}
\usepackage[nolist]{acronym}
\usepackage{tabularx}
\usepackage{mathtools}
\usepackage{amssymb}
\usepackage[tableposition=top]{caption}
\usepackage{structuralanalysis}
\usepackage[para,online,flushleft]{threeparttable}
\usepackage{array}

\def\BibTeX{{\rm B\kern-.05em{\sc i\kern-.025em b}\kern-.08em
    T\kern-.1667em\lower.7ex\hbox{E}\kern-.125emX}}
\usepackage{floatrow}
\usepackage{lipsum}
\usepackage{amssymb}
\usepackage{etoolbox}
\usepackage{float}
\floatstyle{plaintop}
\restylefloat{table}
\makeatletter

\usepackage{MnSymbol}
\newfloatcommand{capbtabbox}{table}[][\FBwidth]

\usepackage{blindtext}

\begin{document}

\title{Big Data-driven Automated Anomaly Detection and Performance Forecasting in Mobile Networks
{\footnotesize \textsuperscript{*}}
\thanks{This project has received funding from the European Union’s Horizon 2020 research and innovation programme under the Marie Sk\l{}odowska-Curie grant agreement No 712949  (TECNIOspring PLUS) and from the Agency for Business Competitiveness of the Government of Catalonia. The work by Mario Garc\'{i}a-Lozano has been funded by the Spanish ministry of science through the project CRIN-5G (RTI2018-099880-B-C32) and with ERFD funds.}
}
\author{\IEEEauthorblockN{\textsuperscript{} Jessica Moysen}
\IEEEauthorblockA{
\textit{Elisa Corporation, Helsinki, Finland }\\
and Fundaci\`{o} i2CAT \\
jessica.moysen@i2cat.net}
\and
\IEEEauthorblockN{\textsuperscript{} Furqan Ahmed}
\IEEEauthorblockA{ 
\textit{Elisa Corporation}\\
Helsinki, Finland \\
furqan.ahmed@elisa.fi}
\and
\IEEEauthorblockN{\textsuperscript{} Mario Garc\'{i}a-Lozano}
\IEEEauthorblockA{
\textit{Universitat Polit\`{e}cnica de Catalunya}\\
Barcelona, Spain \\
mariogarcia@tsc.upc.edu}
\and
\IEEEauthorblockN{\textsuperscript{} Jarno Niemel\"{a}}
\IEEEauthorblockA{
\textit{Elisa Corporation}\\
Helsinki, Finland \\
jarno.niemela@elisa.fi}

}
 \begin{acronym}[AAAAAAA]
\acro{AAS}[AAS]{Active Antenna Systems}
\acro{AC}[AC]{Actor Critic}
\acro{AI}[AI]{ Artificial Intelligence}
\acro{ABS}[ABS]{Almost Blank Subframes}
\acro{AIP}[AIP]{Administrative Incentive Pricing}
\acro{ANA}[ANA]{Autonomic Network Architecture}
\acro{ANR}[ANR]{Automatic Neighbour Relation}
\acro{ANN}[ANN]{Artificial Neural Network}
\acro{AP}[AP]{Access Point}
\acro{API}[API]{Application Programming Interface}
\acro{B5G}[B5G]{beyond 5G}
\acro{BeFemto}[BeFemto]{Broadband Evolved FEMTO Networks}
\acro{BER}[BER]{Bit Error Rate}
\acro{BLER}[BLER]{Block Error Rate}
\acro{BIONETS}[BIONETS]{Genetically inspired networks}
\acro{2TBN}[2TBN]{two-stage temporal Bayesian network}
\acro{BS}[BS]{Base Station}
\acro{CESC}[CESC]{Cloud-Enabled Small Cell}
\acro{COGEU}[COGEU]{Cognitive radio systems for efficient sharing of TV white spaces in EUropean context}
\acro{CAPEX}[CAPEX]{Capital Expenditure}
\acro{CASCADAS}[CASCADAS]{Component-ware for Autonomic, Situation-aware Communications, and Dynamically Adaptable Services}
\acro{CATNETS}[CATNETS]{Evaluation of the Catallaxy Paradigm for Decentralized Operation of Dynamic Application Networks}
\acro{CET}[CET]{Changes Electrical Tilt}
\acro{CG}[CG]{Coordination Game}
\acro{C-SON}[C-SON]{centralized SON}
\acro{C-RRM}[C-RRM]{centralized RRM}
\acro{CIO}[CIO]{Cell Individual Offset}
\acro{CCO}[CCO]{Coverage and Capacity Optimization}
\acro{COR}[COR]{Cell Outage Recovery}
\acro{CDF}[CDF]{Cumulative Distribution Function}
\acro{CDR}[CDR]{Charging Data Records}
\acro{COC}[COC]{Cell Outage Compensation}
\acro{COD}[COD]{Cell Outage Detection}
\acro{COM}[COM]{Cell Outage Management}
\acro{CNN}[CNN]{Convolutional Neural Network}
\acro{LSTM}[LSTM]{Long Short-Term Memory}
\acro{RNN}[RNN]{Recurrent Neural Network}
\acro{CoMP}[CoMP]{Coordinated Multi Points}
\acro{COOPCOM}[COOPCOM]{Comunicaciones Cooperativas y Oportunistas en Redes Inal\'{a}mbricas}
\acro{COGNET}[COGNET]{Cognitive networks}
\acro{CRF}[CRF]{Conditional Random Field}
\acro{CRS}[CRS]{Cognitive Radio System}
\acro{CQI}[CQI]{Channel Quality Indicator}
\acro{CM}[CM]{Configuration Management}
\acro{IM}[IM]{Inventory Management}
\acro{CTTC}[CTTC]{Centre Tecnològic de Telecomunicacions de Catalunya}
\acro{DA}[DA]{Discriminant Analysis}
\acro{D-SON}[D-SON]{distributed SON}
\acro{D-RRM}[D-RRM]{distributed RRM}
\acro{DCI}[DCI]{Data Control Indication}
\acro{DBSCAN}[DBSCAN]{Density-Based Spatial Clustering of Applications with Noise}
\acro{HDBSCAN}[HDBSCAN]{Hierachical-Density-Based Spatial Clustering of Applications with Noise}

\acro{DBM}[DBM]{Deep Boltzmann Machine}
\acro{DBN}[DBN]{Deep Belief Network}
\acro{DNN}[DNN]{Deep Neural Network}
\acro{DL}[DL]{Downlink}
\acro{DP}[DP]{Dynamic Programming}
\acro{DSA}[DSA]{Dynamic Spectrum Access}
\acro{DT}[DT]{Decision Trees}
\acro{E3}[E3]{End-to-End Efficiency}
\acro{eNB}[eNB]{Enhanced Node Base station}
\acro{eNBs}[eNBs]{Enhanced Node Base stations}
\acro{EIRP}[EIRP]{Equivalent Isotropically Radiated Power}
\acro{EPC}[EPC]{Evolved Packet Core}
\acro{ETRI}[ETRI]{Electronics and Telecomunications Research Institute}
\acro{ES}[ES]{Energy Saving}
\acro{E-UTRAN}[E-UTRAN]{Evolved Universal Terrestrial Radio Access Network}
\acro{ETSI}[ETSI]{European Telecommunications Standards Institute}
\acro{FFR}[FFR]{Fractional Frequency Reuse}
\acro{FL}[FL]{Fuzzy Logic}
\acro{FE}[FE]{Feature Extraction}
\acro{FS}[FS]{Feature Selection}
\acro{GT}[GT]{Game Theory}
\acro{Gandalf}[Gandalf]{Monitoring and Self-tuning of RRM parameters in a Multi-System Network}
\acro{GLM}[GLM]{Generalized Linear Models}
\acro{GPI}[GPI]{Generalized Policy Iteration}
\acro{GPRS}[GPRS]{General Packet Radio Service}
\acro{GSM}[GSM]{Global System for Mobile Communications}
\acro{3GPP}[3GPP]{3rd Generation Partnership Project}
\acro{5GPPP}[5GPPP]{5G Infrastructure Public Private Partnership}
\acro{GPRS}[GPRS]{General Packet Radio Service}
\acro{GSM}[GSM]{General System for Mobile Communications}
\acro{HAGGEL}[HAGGEL]{An Innovative Paradigm for Autonomic Opportunistic Communication}
\acro{HeNB}[HeNB]{Home eNodeB}
\acro{Het-Net}[Het-Net]{Heterogenous Network}
\acro{HMM}[HMM]{Hidden Markov Model}
\acro{HO}[HO]{Handover}
\acro{HOF}[HOF]{Handover Failure}
\acro{HII}[HII]{High Interference Indicator}
\acro{IRP}[IRP]{Integration Reference Point}
\acro{IS}[IS]{Information Service}
\acro{IEEE}[IEEE]{Institute of Electrical and Electronics Engineers}
\acro{ICIC}[ICIC]{Inter-Cell Interference Coordination}
\acro{IMS}[IMS]{IP Multimedia Subsystem}
\acro{IoT}[IoT]{Internet of Things}
\acro{IRAT}[IRAT]{Inter-Radio Access Technology}
\acro{$k$-NN}[$k$-NN]{$k$-Nearest Neighbors}
\acro{KPI}[KPI]{Key Performance Indicator}
\acro{LENA}[LENA]{LTE-EPC Network Simulator}
\acro{LTE}[LTE]{Long Term Evolution}
\acro{LTE-Advanced}[LTE-Advanced]{Long Term Evolution Advanced}
\acro{LB}[LB]{Load Balancing}
\acro{LTE-U}[LTE-U]{LTE-Unlicensed}
\acro{LAA}[LAA]{Licensed Assisted Access}
\acro{MAC}[MAC]{Media Access Control}
\acro{MDT}[MDT]{Minimization of Drive Tests}
\acro{M2M}[M2M]{Machine to Machine}
\acro{MIMO}[MIMO]{Multiple-input Multiple-output}
\acro{MC}[MC]{Monte Carlo}
\acro{MCS}[MCS]{Modulation and Coding Scheme}
\acro{MEC}[MEC]{Mobile Edge Computing}
\acro{MDP}[MDP]{Markov Decision Process}
\acro{MME}[MME]{Mobility Management Entity}
\acro{MONOTAS}[MONOTAS]{Mobile Network Optimisation Through Advanced Simulation}
\acro{MCC}[MCC]{Mobile Cloud Computing}
\acro{MLB}[MLB]{Mobility Load Balancing}
\acro{ML}[ML]{Machine Learning}
\acro{MLL}[ML]{Maximum Likelihood}
\acro{MRO}[MRO]{Mobility Robustness$/$Handover Optimisation}
\acro{MWC}[MWC]{Mobile World Congress}
\acro{NB}[NB]{Naive Bayes}
\acro{NBs}[NBs]{Node Base station}
\acro{NE}[NE]{Network Element}
\acro{NET-REFOUND}[NET-REFOUND]{Network research foundations}
\acro{NF}[NF]{Network Function}
\acro{NGMN}[NGMN]{Next Generation Mobile Networks}
\acro{NMS}[NMS]{Network Management Systems}
\acro{NM}[NM]{Network Management}
\acro{NRM}[NRM]{Network Resource Model}
\acro{NFV}[NFV]{Network Functions Virtualisation}
\acro{NFVI}[NFVI]{NFV Infraestructure}
\acro{OFDMA}[OFDMA]{Orthogonal Frequency Division Multiple Access}
\acro{OI}[OI]{Overload Indicator}
\acro{OSS}[OSS]{Operation and Support System}
\acro{OPEX}[OPEX]{Operational Expenditure}
\acro{OAM}[OAM]{Operation and Maintenance Center}
\acro{PCI}[PCI]{Physical Cell ID}
\acro{PC}[PC]{Principal Component}
\acro{Pc}[Pc]{Power control}
\acro{PCA}[PCA]{Principal component analysis}
\acro{PCI}[PCI]{Automated Configuration of Physical Cell Identity}
\acro{PItoRC}[PItoRC]{Policy Iteration to Resource Conflicts}
\acro{PDF}[PDF]{Probability Density Function}
\acro{PGW}[PGW]{PDN Gateway}
\acro{PDSCH}[PDSCH]{Physical Downlink Shared Channel}
\acro{PDU}[PDU]{Protocol Data Unit}
\acro{PDSCH}[PDSCH]{Physical Downlink Shared Channel}
\acro{PUSCH}[PUSCH]{Physical Uplink Shared Channel}
\acro{PDCCH}[PDCCH]{Physical Downlink Control Channel}
\acro{PGW}[PGW]{PDN Gateway}
\acro{SGW}[SGW]{Serving Gateway}
\acro{PF}[PF]{Proportional Fair Scheduler}
\acro{PHR}[PHR]{Power Headroom Report}
\acro{PM}[PM]{Performance Management}
\acro{PRB}[PRB]{Physical Resource Block}
\acro{PS}[PS]{Packet Switched}
\acro{PSD}[PSD]{Power Spectral Density}
\acro{PU}[PU]{Primary User}
\acro{QoS}[QoS]{Quality of Service}
\acro{QoE}[QoE]{Quality of Experience}
\acro{QAM}[QAM]{Quadrature Amplitude Modulation}
\acro{RBF}[RBF]{Radial basis function}
\acro{RBM}[RBM]{Restricted Botlzmann Machine}
\acro{RACH}[RACH]{Random Access Channel}
\acro{RAN}[RAN]{Radio Access Network}
\acro{RAT}[RAT]{Radio Access Technologies}
\acro{RET}[RET]{Remote Electrical Tilt}
\acro{RF}[RF]{Random Forest}
\acro{RB}[RB]{Resource Block}
\acro{RBG}[RBG]{Resource Block Group}
\acro{RBG}[RBG]{Restricted Boltzmann Machine}
\acro{REM}[REM]{Radio Environment Map}
\acro{RL}[RL]{Reinforcement Learning}
\acro{RLF}[RLF]{Radio Link Failure}
\acro{RNTP}[RNTP]{Relative Narrowband Transmit Power}
\acro{RLC}[RLC]{Radio Link Control}
\acro{RMSE}[RMSE]{Root Mean Squared Error}
\acro{RRC}[RRC]{Radio Resource Control}
\acro{RRM}[RRM]{Radio Resource Management}
\acro{RS}[RS]{Reference Signal}
\acro{RSRP}[RSRP]{Reference Symbol Received Power}
\acro{RSRQ}[RSRQ]{Reference Symbol Received Quality}
\acro{SAC}[SAC]{Situated Autonomic Communications}
\acro{SDSE}[SDN]{Self-Organized Network}
\acro{SDSE}[SDSE]{Strongly Dominant Strategy Equilibrium}
\acro{SELFNET}[SELFNET]{Framework for Self-Organized Network Management in virtualized
and Software Defined Networks}
\acro{SEMAFOUR}[SEMAFOUR]{Self-Management for Unified Heterogeneous Radio Access Networks}
\acro{SESAME}[SESAME]{Small cell coordination for multi-tenancy and edge services}
\acro{SG}[SG]{Stochastic Games}
\acro{SGSN}[SGSN]{Serving GPRS Support Node}
\acro{SH}[SH]{Self Healing}
\acro{SINR}[SINR]{Signal to Interference and Noise Ratio}
\acro{SLA}[SLA]{Service Level Agreement}
\acro{SM}[SM]{Saturation Mode}
\acro{SML}[SML]{Stochastic Maximum Likelihood}
\acro{SPCA}[SPCA]{Sparse Principal Component Analysis}
\acro{SVM}[SVMs]{Support Vector Machines}
\acro{SVR}[SVR]{Support Vector Regression}
\acro{SDN}[SDN]{Software Defined Network}
\acro{TCE}[TCE]{Trace Collection Entity}
\acro{SL}[SL]{Supervised Learning}
\acro{SO}[SO]{Self organization}
\acro{SOCRATES}[SOCRATES]{Self-Optimisation and self-ConfiguRATion in wirelEss networkS}
\acro{SOFOCLES}[SOFOCLES]{Self-organized FemtOCeLls for broadband sErviceS}
\acro{SOM}[SOM]{Self-organizing Map}
\acro{SON}[SON]{Self-organizing Networks}
\acro{SOS}[SOS]{Self organized System}
\acro{SS}[SS]{Solution Sets}
\acro{SL}[SL]{Supervised Learning}
\acro{SIMO}[SIMO]{Single-input Multiple-output}
\acro{subMDP}[subMDP]{Markov decision sub-process}
\acro{TA}[TA]{Timing Advance}
\acro{TCE}[TCE]{Trace Collection Entity}
\acro{TCP}[TCP]{Transmission Control Protocol}
\acro{TD}[TD]{Temporal Difference}
\acro{TTT}[TTT]{Time to Trigger}
\acro{TTI}[TTI]{Transmission Time Interval}
\acro{TBS}[TBS]{Transport Block Size}
\acro{TXP}[TXP]{Transmission Power}
\acro{UDN}[UDN]{Ultra ­Dense Network}
\acro{UE}[UE]{User Equipment}
\acro{UEs}[UEs]{User Equipments}
\acro{UMTS}[UMTS]{Universal Mobile Telecommunications System}
\acro{UL}[UL]{Unsupervised Learning}
\acro{UTRAN}[UTRAN]{Universal Terrestrial Radio Access Network}
\acro{WLAN}[WLAN]{Wireless Local Area Network}
\acro{WCQI}[WCQI]{wideband CQI}
\acro{AMF}[AMF]{Access and Mobility Function}
\acro{UPF}[UPF]{User Plane Function}
\acro{5G}[5G]{5th generation}
\acro{5GC}[5GC]{Fifth Generation Core Network}
\acro{NG-RAN}[NG-RAN]{Next Generation RAN}
\acro{NG-SON}[NG-RAN]{Next Generation SON}
\acro{NRT}[NRT]{Neighbor Relation Table}
\acro{NR}[NR]{New Radio}
\acro{IT}[IT]{Information Technology}
\acro{COTS}[COTS]{Commercial off-the-shelf}
\acro{VNN}[VNN]{vehicular nomadic nodes}
\acro{V-SON}[V-SON]{virtualized SON}
\acro{VNF}[VNF]{Virtual Network Function}
\acro{ONF}[ONF]{Open Networking Foundation}
\acro{URLLC}[URLLC]{ultra-reliable and low-latency communications}
\acro{mMTC}[mMTC]{massive machine-type communications}
\acro{eMBB}[eMBB]{enhanced mobile broadband}
\acro{RNTI}[RNTI]{Radio Network Temporary Identifier}
\acro{IMSI}[IMSI]{International Mobile Subscriber Identity}
\acro{TB}[TB]{Transport Block}
\acro{PDCP}[PDCP]{Packet Data Convergence Protocol}
\acro{PHY}[PHY]{Physical layer}
\acro{EOS}[EOS]{Experience Oriented SON}
\end{acronym} 
\maketitle

\begin{abstract}
The massive amount of data available in operational mobile networks offers an invaluable opportunity for operators to detect and analyze possible anomalies and predict network performance. In particular, application of advanced machine learning (ML) techniques on data aggregated from multiple sources can lead to important insights, not only for the detection of anomalous behavior but also for performance forecasting, thereby complementing classic network operation and maintenance solutions with intelligent monitoring tools. In this paper, we propose a novel framework that aggregates diverse data sets (e.g. configuration, performance, inventory, locations, user speeds) from an operational LTE network and applies ML algorithms to diagnose network issues and analyze their impact on key performance indicators. To this end, pattern identification and time-series forecasting algorithms are used on the ingested data. Results show that proposed framework can indeed be leveraged to automate the identification of anomalous behaviors associated with the spatial-temporal characteristics, and predict customer impact in an accurate manner.
\end{abstract}

\begin{IEEEkeywords}
Mobile networks, network management, big data, machine learning, anomaly detection, algorithms
\end{IEEEkeywords}

\IEEEpeerreviewmaketitle 

\section{Introduction}
\label{sec:intro}
The 5th generation (5G) networks differ substantially from the previous generations, leading to new challenges and opportunities. Most importantly, 5G must be able to support new types of frequency bands, multi-connectivity, spectrum and technologies. To fulfill these requirements, automation is becoming increasingly important. Increased operation and management complexity calls for novel intelligent solutions capable of not only handling the growing complexity, heterogeneity, and dynamic nature of the network, but also the monitoring of massive amounts of performance data and detection of anomalous patterns in it.
Another key driving factor is the challenging requirements of 5G and beyond 5G networks, which differ from \ac{LTE} in a multitude of ways. For example, the first version of LTE (Release 8) has about 8$\times$ less layer 1 parameters to be controlled via \ac{RRC}, when compared to the first version of New Radio (Release 15). Therefore, intelligent automation is of paramount importance for the network management, optimization, and monitoring. Unsurprisingly, operators and vendors have tremendous interest in evaluating the potential benefits of automation and related challenges that future mobile networks will face as advanced features of 5G networks are rolled-out\cite{saad2019vision, tariq2019speculative}.

Recent advances in the field of cloud technologies and data processing capabilities have paved the way for powering mobile network automation solutions using machine learning (ML) techniques \cite{challita, jmoysenELSEVIER}. In regards to this, a key issue is the efficient processing of network data characterized by challenges related to data volume, variety, velocity, and veracity. In the context of
mobile networks, data with such properties is referred to as \emph{big data} \cite{imran_bson}. In the context of mobile network automation, ML in conjunction with big data is emerging as a potent tool for optimizing increasingly complex networks\cite{joseph_mro, imran_bson}. Moreover, apart from network optimization, it can be used for enabling data-driven intelligence for performance monitoring in mobile networks\cite{9064292,7984760,7389830,8108601}. %
Existing work on the monitoring and diagnosis of network faults using ML techniques includes pattern identification, grouping, and learning\cite{rcabarco,ulbarco}. This approach is motivated by the fact that massive amounts of data generated in mobile networks necessitates pre-processing tasks such as data reduction using feature selection and/or feature extraction techniques. In \cite{rcabarco}, the authors propose a framework for managing high-dimensional data and optimizing the performance of root cause analysis, whereas \cite{ulbarco} proposes an automatic diagnosis system that takes advantage of self-organizing maps to build a system capable of identifying anomalous behaviors in networks. Likewise, the approaches discussed for mobile traffic forecasting in \cite{kaisa} and \cite{xu} are inspired by the estimation problem. In \cite{kaisa}, causal analysis and long short term memory (LSTM) models are used to predict call detailed records. On the other hand, in \cite{xu}, the authors focus on extracting and modeling traffic patterns of base stations in a network. Based on a time series analysis approach, mobile traffic is decomposed into regularity and randomness components, which is subsequently used to forecast the traffic patterns.

\begin{figure*}[t!]
\centering
\includegraphics[trim={0.0cm 2cm 0.0cm 1.95cm},scale=0.45,clip]{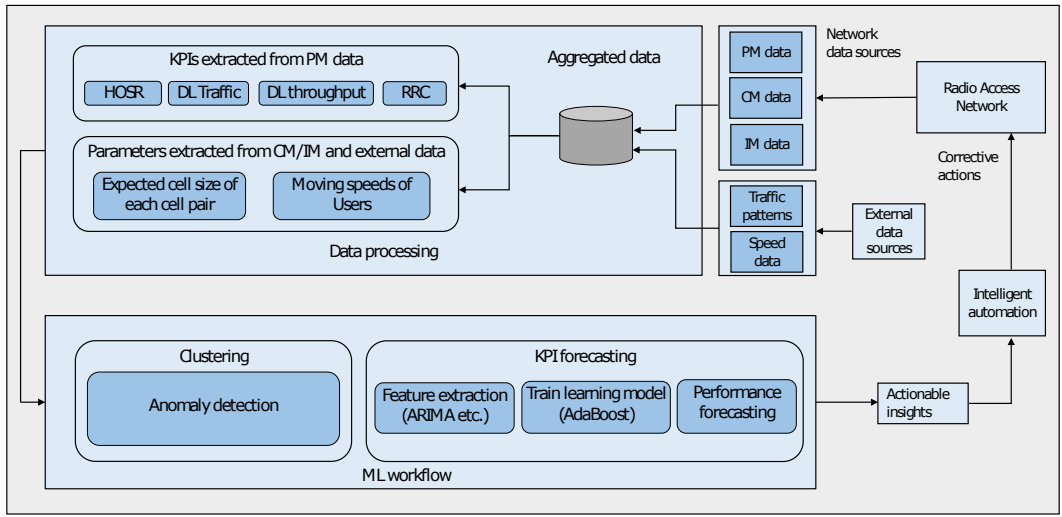}
\caption{Big data-driven automated anomaly detection and prediction framework.}
\label{fig1}
\end{figure*}
In this paper, we propose a framework for analyzing possible anomalies and monitoring mobile network performance using data aggregated from a variety of data sources. The proposed framework: (1) collects the most representative key performance indications (KPIs) from the data ingested from a live mobile network; (2) constructs time series and applies clustering to classify them based on geographical characteristics; and (3) makes use of feature extraction techniques and regression models for performance forecasting. 
The approach is based on identifying patterns in the spatio-temporal signatures of cells. The capability of predicting the performances based on the grouping patterns identified using clustering paves the way for designing fine grained mitigating actions for each individual group of cells.

The rest of the paper is organized as follows. Section~\ref{sec:framework} describes the proposed framework including data processing and ML workflow used for its processing and analysis. Section~\ref{sec:implementation} presents the details regarding implementation of the propose framework and results. Finally, we draw our conclusions and discuss possible directions for future work in Section~\ref{sec:con}.

\section{Framework}
\label{sec:framework}
The proposed framework comprises of two main steps namely input data processing and ML workflow. Data processing involves ingestion of mobile network data from different sources and its aggregation, followed by clustering and KPI forecasting which constitutes ML workflow step as depicted in Fig.~\ref{fig1}. Next, we discuss these steps in more detail.  
\subsection{Data Processing}
The main challenges involved in mobile network data ingestion, cleaning, and aggregation, stem from the challenging characteristics of the data sets in terms of volume, velocity, variety, and veracity. For example, inventory management (IM) data consists of equipment information, which is almost static. On the other hand, performance management (PM) data involves thousands of counters that collect raw measurements from that network. Thus, it grows continuously and requires time-series database for efficient storage and querying. Long term storage is particularly challenging, but nevertheless inevitable for some ML based use-cases. In this work, we consider following mobile network data sources.

\subsubsection{CM and IM Data} 
\ac{CM} data set essentially consists of radio access network configuration parameters such eNB/cell ID, frequency bands, neighbor relations, and settings of various features etc. These parameters determine the performance of the network, and are adjusted during the roll-out and optimization activities. On the contrary, IM data consists of information related to location and inventory items such as equipment (base station, antenna types etc.), services, and other infrastructure. 

\subsubsection{PM Data}
\ac{PM} data comprises of performance counters which collect different measurements at regular intervals with a very high granularity (e.g. 15 minutes to one hour), resulting in massive amounts of time-series data. Moreover, measurements are collected at various levels e.g. cell, neighbor relation etc. For example, cell throughput counter collects cell level throughput, whereas handover counters collect handover statistics per neighbor relation. However, PM data is used not only in performance monitoring but also in the ML based prediction algorithms. 

\subsubsection{Movement and Speed Data}
Mobility characteristics such as movement patterns and speed of the mobile users are very important for predicting network performance from a user perspective. Here, we use Google Maps platform to estimate user's speed at a given location. By using this information in conjunction with the cell location, it is possible to estimate the speed of users in a cluster of cells located in a certain area. 

\subsection{Workflow}
The ML workflow comprises of two main steps, namely \emph{clustering} and \emph{time series forecasting}. As shown in Fig.~\ref{fig1}, the first step creates a group of clusters based on spatial characteristics e.g. cell location and estimated user's speed, followed by the detection of anomalous behaviours associated with temporal characteristics. In the second step, features are extracted and network performances is predicted using time-series forecasting to enable proactive troubleshooting in the cases where anomalous behavior is expected.

\subsubsection{Clustering}
\label{cluster}
The idea is to create of groups of signatures to understand the spatio-temporal characteristics and how they correlate with the performance experienced by end users. To this end, the first step creates clusters using data on cell location, estimated user speeds.
The PM data considered includes different cell KPIs related to traffic, congestion, and mobility problem activities during a number of hours. The number of cells and duration of the raw measurement intervals determines the granularity of the data. Let us consider a cell~$i \in \mathcal{I}$, where the number of cells in the whole network is~$N$. Let $ \mathbf{x}_i = \left[x_{i,1} \dots x_{i,m} \right]
^{T}\in \mathbb{R}^{m}$ denote the vector of KPI values for cell $i$. The data for all the cells is collected in a matrix $\mathbf{X} = [ \mathbf{x}_1,  \dots, \mathbf{x}_{N}] \in \mathbb{R}^{{N} \times m}$.  
Once the time series is created, signatures belonging to same geographical regions are grouped together. We consider geographical information for partitioning $\mathbf{X}$, the resulting partition $\mathcal{C}$ comprises of $K$ labelled clusters $\mathcal{C} \coloneqq \left(C_1, \dots C_k \dots C_K \right)$. That is, the output is a set of classes of signatures, denoted as $\mathcal{C}$.
We aggregate statistics of each KPI considered for the cells in each cluster. The location and velocities of users present in the coverage area of each cell is also taken into account. The output is a correlation analysis that allows us to detect if a cluster is associated to a particular set of anomalies. By analysing each group of signatures over time, anomalous behaviour impacting network performances can be identified.

\subsubsection{KPI forecasting}
\label{tsf}
The next step is to predict how the considered KPIs will evolve (e.g. when and in which cluster anomalous behavior is likely to occur). This gives a dynamic picture of the network state, including information that can be used to handle expected performance degradation in a preemptive manner. 
Following the approach in \cite{tsfresh}, an efficient reduction of time-series data is done first via feature extraction process, where each feature captures a specific measurable characteristic of the time-series. The statistical features considered are derived from basic summary statistics and aspects of sample distribution. Moreover, features from observed dynamics are also considered. These include Auto-regressive integrated moving average (ARIMA) model coefficients obtained by fitting the unconditional maximum likelihood of an autoregressive process on the time-series with a maximum lag $m_{\rm T}$, mean absolute change, and mean auto-correlation. A full summary of the features is delineated in TABLE~\ref{tab2}.  
\begin{table}[t!]
\centering
\caption{Extracted features}
\label{tab2}
\footnotesize\begin{tabular}[width=\columnwidth]{p{25ex}p{38ex}}
\hline
\textbf{From sample statistics} &  \textbf{Description} \\
\hline
\hline Maximum($\mathbf{x}$)& Maximum sample of time series $\mathbf{x}$.\\
\hline Minimum($\mathbf{x}$) & Minimum sample time series $\mathbf{x}$.\\
\hline Mean($\mathbf{x}$) & Arithmetic mean of time series $\mathbf{x}$. \\
\hline Var($\mathbf{x}$)& Expectation of the squared deviation of time series $\mathbf{x}$ from its mean $\mathbf{X}$ without bias. \\
\hline Skewness($\mathbf{x}$)& Sample calculated with adjusted Fisher-Pearson standardized moment coefficient.\\
\hline Kurtosis($\mathbf{x}$) & Fourth central moment of time series $\mathbf{x}$ divided by the square of its variance \\
\hline Median($\mathbf{x}$) & For a time series $\mathbf{x}$ with an uneven number of samples, the median is the middle of the sorted time series values. \\
\hline
\textbf{From sample distribution} &  \textbf{Description} \\
\hline
\hline Variance greater than std($\mathbf{x}$) & The feature indicates if the variance is greater than the standard deviation. \\
\hline Number of values that are above/below median & Number of values, which are larger/lower than the median value of the time series sample \\
\hline
\textbf{From observed dynamics} &  \textbf{Description} \\
\hline
\hline ARIMA model coefficients & The feature meets the unconditional maximum likelihood of an auto-regressive process with a maximum lag $m_{\rm T}$.  \\
\hline Mean absolute change($\mathbf{x}$)& Arithmetic mean of absolute differences between subsequent time series values.\\
\hline Mean auto-correlation($\mathbf{x}$)& Average auto-correlation over possible lags $l$ ranging from $1$ to $m$.\\
\hline
\end{tabular}
\end{table} 
The new features are used to train an ensemble method to predict the next time step. As highlighted in \cite{jmoysenISCC}, ensemble methods are learning models, which combine the opinions of multiple learners\cite{ensamble}. The learning algorithm is run several times, each one with different subset of training samples. We use Ada-Boost regressor, which maintains a set of weights over the original training set, and adjusts these weights by increasing the weight of examples that are miss-classified, and decreasing the weight of examples that are correctly classified. For every value, the historical data is used to fit the model and predict the next value. We evaluate the performance by via a comparison of predicted value against the true value in terms of mean absolute error (MAE), which measures the error between the predicted and true value in terms of the arithmetic average of the absolute errors $|e_i|=|\hat{y}_i- y_i|$, where $\hat{y}_i$ is the prediction and $y_i$ is the true value. 

\section{Implementation \& Results} 
\label{sec:implementation}
For implementation, we use commercial LTE network datasets from a major Finnish mobile operator. The total number of cells in the dataset is $N$ = $5672$, with one week of PM data $m$ = $168$ (i.e granularity of sixty minutes). Following four cell level KPIs described in TABLE~\ref{tab3} are considered to construct time-series matrices $\mathbf{X_1}, \mathbf{X_2}, \mathbf{X_3}, \mathbf{X_4}$: handover success rate (HOSR), downlink throughput, downlink traffic, and RRC connection request rate. 

\begin{figure}[t!]
\centering
\includegraphics[scale=.28]{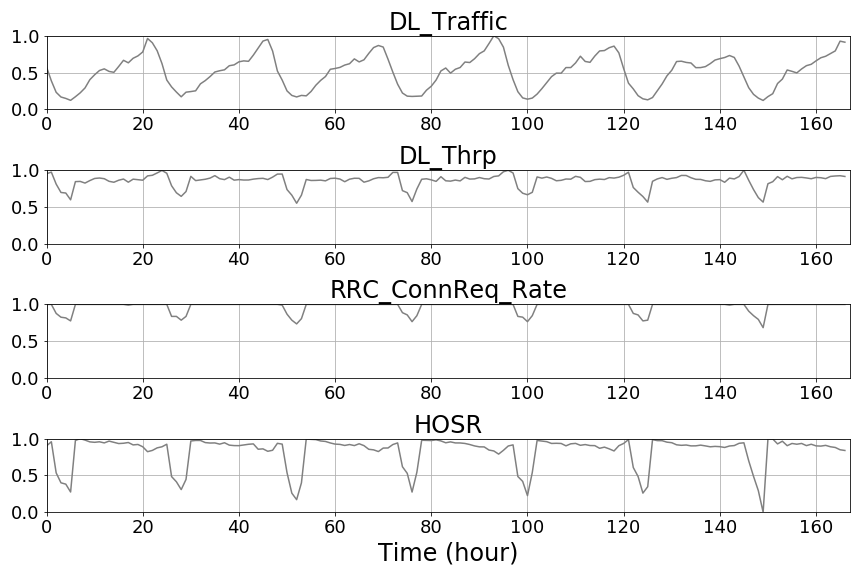}  
   \caption{KPI signatures representing normal behaviours (average speed $> 50$ Km/h).}
\label{fig2}
\end{figure}

\begin{figure}[t]
\centering
\includegraphics[scale=.28]{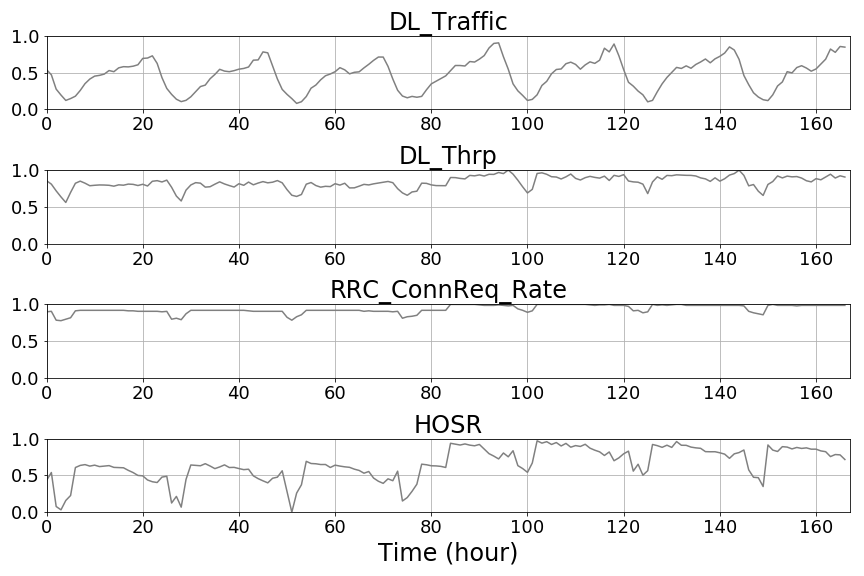} 
\caption{KPI signatures representing abnormal behaviours (average speed $> 50$ Km/h).}
\label{fig3}
\end{figure}

The CM and IM data used for implementing spatial clustering includes cell ID, latitude, longitude, name of the site and eNB, value of bearing(s) and neighbouring cells. The expected cell size of each cell as the $90$-th percentile distance from source cell to target cell. The data related to users' locations and speed measurements comprising of actual traces is downloaded from the Google Maps APIs. This API allows us to get information such as speed limits (from $0$ to $100$ Km/h) and the way points of different travel modes (i.e. driving, walking, cycling and public transportation), which is used to estimate the speed of users likely present in the coverage area. 
Spatial clustering yields a set of $11$ clusters in the network with different speeds and locations. The matrices $\mathbf{X_1}, \mathbf{X_2}, \mathbf{X_3}, \mathbf{X_4}$ are split accordingly for the detection of anomalies in the resulting time-series, at the level of individual clusters. The cluster level time series analysis helps to identify the signatures of mobility related anomalies. For instance, results for a cluster where the average speed is greater than $50$ Km/h (i.e. travel mode of users is driving and$/$or public transportation) are shown in Figures~\ref{fig2} and \ref{fig3}, for normal and the abnormal KPI signatures respectively. The x-axis represents the number of hours in one week, while the y-axis represents the normalised value of each KPI considered. From this figure, we can observe the trend and seasonality KPI variations at the daily and weekly level. Figure~\ref{fig3} suggests that on the first three days of the week (Monday, Tuesday, and Wednesday) the trend component deviates from the normal signature, and is clearly identified as an anomaly. 
\begin{table}[t!]
\centering
\caption{KPIs analyzed from PM data}
\label{tab3}
\footnotesize\begin{tabular}[width=\columnwidth]{p{25ex}p{38ex}}
\hline

\textbf{KPI} &  \textbf{Description} \\
\hline
\hline Handover successful rate (HOSR) percentage &This indicator is calculated as the ratio between successful handovers divided by the total number of handovers every hour.\\
\hline Downlink cell throughput (Mbps)& This indicator is calculated as the downlink throughput at the \ac{PDCP} layer in cell $i$ divided by total duration for transmitting downlink data at the PDCP layer $/$ 1000.\\
\hline Downlink traffic (MB) & This indicator is calculated by the downlink throughput at the \ac{PDCP} layer. \\
\hline \ac{RRC} Connection Request rate percentage & This indicator is calculated as the success ratio for the RRC connection establishment divided by RRC connection request attempts. \\
\hline
\end{tabular}
\end{table} 
\begin{figure}[t]
\centering
\includegraphics[width=\columnwidth]{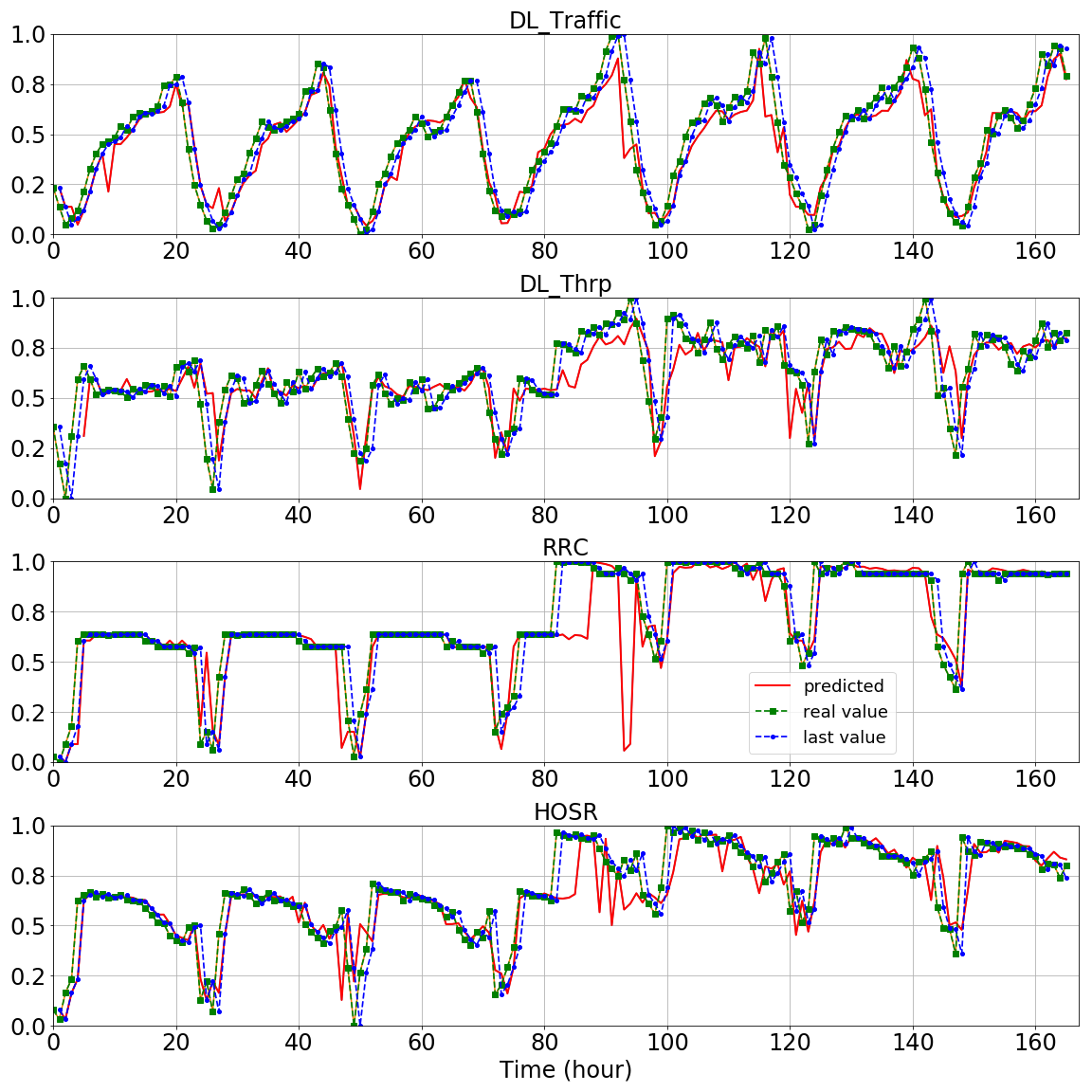}
   \caption{Predicted, last, and real KPI value. Time series with abnormal behaviours.}
\label{fig4a}
\end{figure}
\begin{figure}[t]
  \centering
\includegraphics[width=\columnwidth]{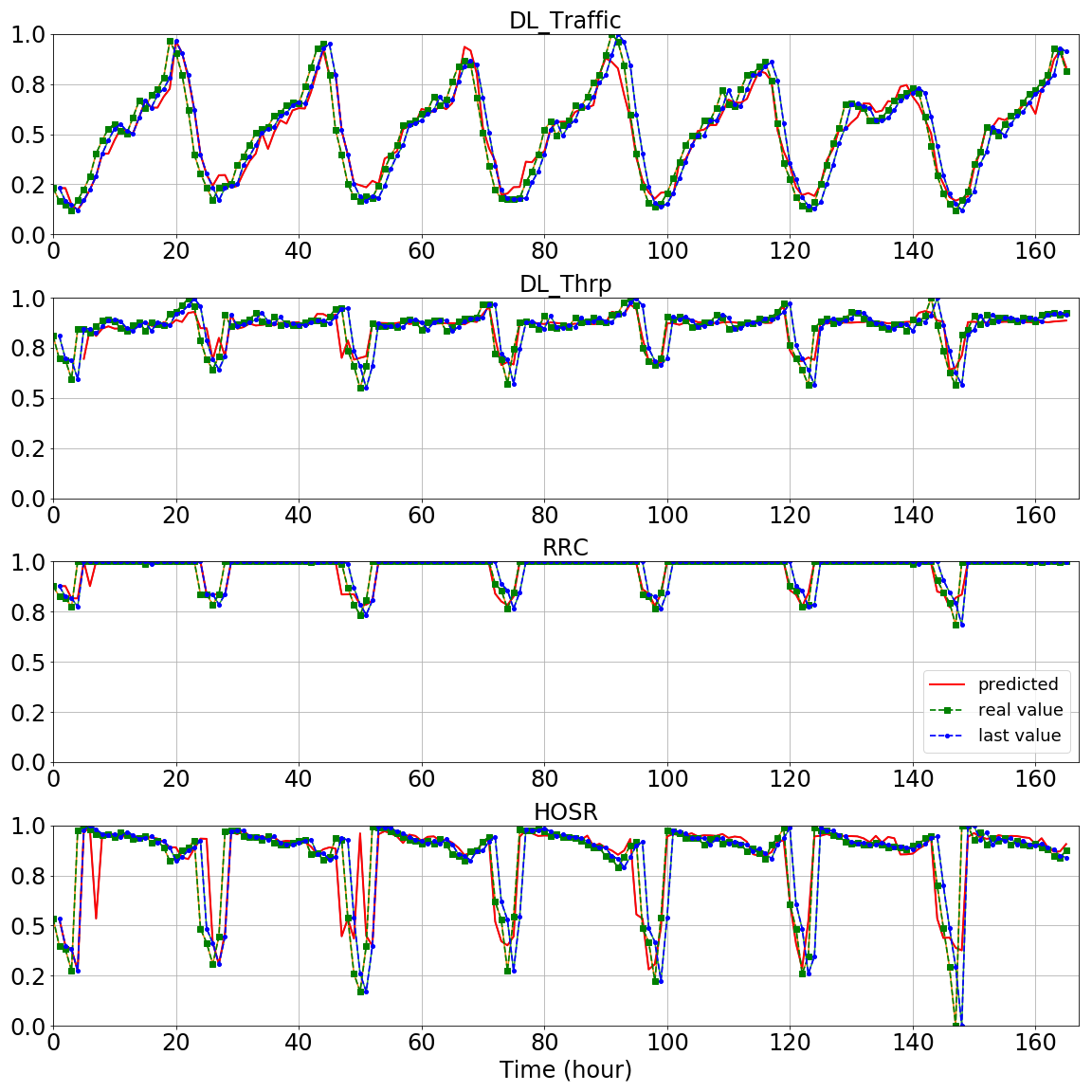}
\caption{Predicted, last, and real KPI value. Time series with normal behaviours.}
  \label{fig4b}
\label{fig4}
\end{figure}
\begin{figure}[t]
\begin{subfigure}{0.49\textwidth}
\centering
\includegraphics[width=\columnwidth]{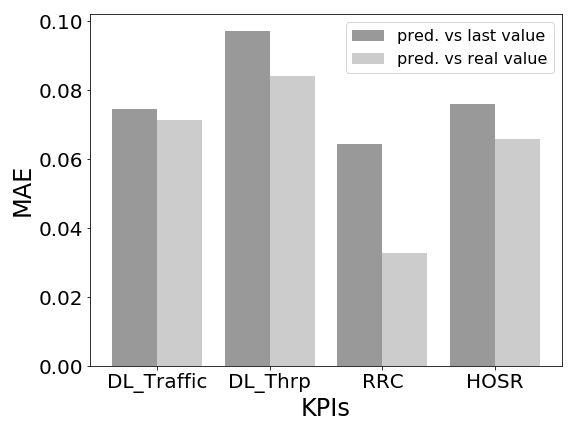}  
   \caption{}
\label{fig5a}
\end{subfigure}
\hfill
\begin{subfigure}{0.49\textwidth}
  \centering
\includegraphics[width=\columnwidth]{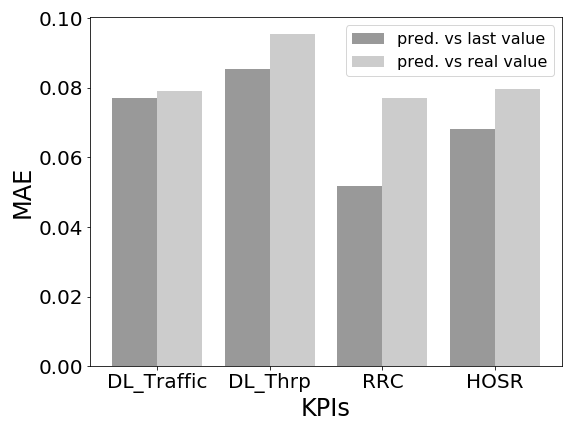}  
\caption{}
 \label{fig5b}
\end{subfigure}
\caption{Mean absolute error considering time series with normal (Fig.~\ref{fig5a}) and abnormal behaviours (Fig.~\ref{fig5b}).}
 \label{fig5}
\end{figure}
Following this procedure, anomalies can be localized and detected at cluster level. The first step towards predicting KPI values is the extraction of features in TABLE~\ref{tab2} from the data. We consider a number of basic features from sample statistics (e.g. maximum, minimum, mean, variance, skewness, kurtosis, and median) and sample distribution (e.g. variance greater than standard deviation, and number of values that are above/below median). From observed dynamics, ARIMA model coefficients with the maximum lag of $m_{\rm T}=20$, mean absolute change, and mean auto-correlation are extracted. For the detection, we use Ada-Boost regressor with a linear loss function. The parameters include learning rate and the number of estimators, which are set to the default values of one and ten respectively.
In the plots shown in Figures~\ref{fig4a} and \ref{fig4b}, the red curve is the output of the Ada-Boost regressor, the green curve is the true value, and as a benchmark we consider the last value before the prediction (blue curve). Figure~\ref{fig4a} shows the performance on the time series with anomalous behaviours, whereas Fig.~\ref{fig4b} corresponds to time series with normal or expected behaviour. From these figures, we observe that both approaches perform quite well. However, Ada-Boost outperforms the benchmark approach on the time series with normal behaviours. This can also be observed in Fig.~\ref{fig5}, which shows the MAE performance metric for each KPI. The dark grey bar corresponds to the error between Ada-Boost regressor (labeled as \emph{predicted}) and benchmark value (labeled as \emph{last value}), whereas the light grey bar corresponds to the error between predicted and real value. It is clear that Ada-Boost regressor outperforms the benchmark approach, as it leads to a smaller error. However, opposite behavior can be observed in Fig.~\ref{fig5b}, where the error between predicted and real value is greater than the error between the predicted value and benchmark last value.
It is worth noting that since historical values are used to generate new features, the accuracy of the Ada-Boost regressor model decreases if anomalies are presented in the data. In such case, considering only the last value for the prediction gives us better results.
The insights thus gained from different KPIs can be used to not  only analyze the performance in detail but also to identify the root-cause and nature of the anomalies, and suggest corrective actions. 

\section{Conclusion}
\label{sec:con}
We have proposed a framework that makes use of various mobile network data sets supplemented by external data sources, for the automated analysis and prediction of mobile network KPIs. The framework is implemented by using some of the common KPIs used by radio network engineers for the diagnosis and troubleshooting tasks. Results demonstrate that clustering based KPI analysis and anomaly detection, and the use of feature extraction and time-series prediction models enables efficient analysis of network data and helps in identifying possible anomalous behavior. This paves the way for improved network performance and operational efficiency, thereby reducing the time that engineers spend analysing huge amount of network performance data. Possible directions for future work include automated root cause analysis of the detected anomalies followed by intelligent and pro-active self-healing mechanisms.

\bibliographystyle{IEEEtran}
\bibliography{mybibtex}

\end{document}